\documentclass[a4paper,11pt]{article}
\pdfoutput=1 

\usepackage{jheppub} 

\usepackage[T1]{fontenc} 
\usepackage[makeroom]{cancel}
\usepackage{physics}
\usepackage{empheq}
\usepackage{graphicx}
\usepackage{subcaption}
\usepackage{tabularx}
\usepackage{natbib}

\usepackage{amsmath}
\usepackage{amssymb}
\usepackage{hyperref}
\usepackage{enumitem}
\usepackage[utf8]{inputenc}
\usepackage[english]{babel}
\usepackage{xcolor}
\usepackage{soul}
\sethlcolor{yellow}
\usepackage[normalem]{ulem}

\newcommand{\eq}[1]{(\ref{#1})}

\definecolor{blue}{rgb}{0.0, 0.0, 1.0}
\newcommand{\red}[1]{\textcolor{red}{#1}}

\newcommand{\ba}{\begin{eqnarray}}
\newcommand{\ea}{\end{eqnarray}}
\newcommand{\nn}{\nonumber}
\def \be {\begin{equation}}
\def \ee {\end{equation}}
\def \bea {\begin{eqnarray}}
\def \eea {\end{eqnarray}}

\title{\boldmath Black Hole Shadow with Soft Hairs}

\author{Feng-Li Lin, Avani Patel, Hung-Yi Pu}

\affiliation{Department of Physics, National Taiwan Normal University, Taipei, Taiwan 116}
\affiliation{Center of Astronomy and Gravitation, National Taiwan Normal University, Taipei 116, Taiwan}

\emailAdd{linfl@gapps.ntnu.edu.tw, avani.physics@gmail.com, hypu@gapps.ntnu.edu.tw}

\abstract{Light bending by the strong gravity around the black hole will form  the so-called black hole shadow, the shape of which can shed light on the structure of the near-horizon geometry to possibly reveal novel physics of strong gravity and black hole. In this work, we adopt both analytical and ray-tracing methods to study the black hole shadow in the presence of the infrared structure of gravity theory, which manifests the asymptotic symmetries of spacetime as the supertranslation soft hairs of the black hole. Though the black hole metrics with and without the soft hair are related by large gauge transformations, the near horizon geometries relevant for the shape of the shadow are quite different. Moreover, the Hamiltonian for the geodesic seems intrinsically different, i.e., the loss of separability due to the breaking of spherical symmetry by soft hair. By applying ray-tracing computations, we find that the soft hair, although not affecting the shape of the shadow, may change the average size and position of the shadow. Images resulting from soft hair black holes with surrounding accretion flows are also discussed. 
}

\begin{document} 
\maketitle
\flushbottom

\section{Introduction}

Light bending around a heavy object such as the Sun has been a direct experimental confirmation of the theory of general relativity, and is coined as the name of ``gravitational lensing". Moreover, a detailed study of the lensing pattern can reveal the nature of the astrophysical compact objects such as the mysterious black hole, which is thought of as the vacuum solution of Einstein gravity, or their surrounding environment such as the accretion disk. With the successful observation of the the first image of the supermassive black hole located at the center of the galaxy M87 via  Event Horizon Telescope (EHT) \cite{Akiyama:2019cqa,Akiyama:2019brx,Akiyama:2019sww,Akiyama:2019bqs,Akiyama:2019fyp,Akiyama:2019eap}, a global radio observation network, electromagnetic investigation of black hole spacetime in strong gravity regime is now achievable. The light rays coming from uniformly distributed luminous objects in the universe reach the observer from all directions. If there is a black hole between the observer and the source, then the light ray will bend due to the gravity of the black hole. All the light rays which come under the effect of the black hole's gravitational field can be classified into two classes. 
If we trace the light rays back from the point of observation, then the rays corresponding to the first class will end up at a light source, while those corresponding to the second class will hit the horizon \cite{Perlick:2021aok}. The shadow of a black hole consists of a dark part in the observer's sky, which corresponds to the light-rays of the second class.  More specifically, from the observer's image plane, the boundary of the black hole shadow is defined by the division of the observed null geodesics which neither flies to infinity nor goes to the horizon backward in time. The boundary of the shadow are contributed by photons which was tangential to unstable photon orbits with constant radius. For Schwarzschild black hole, the photon orbits has the same radius. The collection of all these possible photon orbits\footnote{The orbits may called "light rings", while a similar name "photon ring" may refer to other meanings in different literature. See \cite{Perlick:2021aok} and discussions therein.} forms the "photon sphere". As solely determined by the backgroud metric, the shadow cast by the black hole provides a unique opportunity to test General Relativity(GR) in the regime of strong gravity \cite{Perlick:2021aok}. For example, near-horizon properties of physical charges predicted by various black hole solutions within GR and beyond can be constrained, via the shadow sizes of different black hole solutions \cite{EventHorizonTelescope:2021dqv}.

The properties of the shadow and its size represent valuable observable common to all metric theories of gravity and can be used to test them for their agreement with EHT measurements, see \cite{Cunha:2019hzj,Li:2020drn,Liu:2020ola,Chen:2020aix,Herdeiro:2021lwl,kon16,you18,Konoplya:2021slg,Li:2021mnx,Kumar:2020owy,Moura:2021eln,Zhang:2021pvx,Kasuya:2021cpk,Addazi:2021pty,Sarkar:2021djs} for the recent progress on such investigations for various geometries of black hole or black hole alternative. Along this line of research, we consider an important class of black holes within the context of GR, which are black holes carrying the soft hairs associated with the asymptotic symmetries of spacetime. These asymptotic symmetries are the manifest of the infrared properties due to long-ranged features of the gravity force such as the loop infrared divergence and its cancellation by soft emission \cite{Weinberg:1965nx,Chung:1965zza}, and are supposed to yield the gravitational memory effect \cite{1987Natur.327..123B,Thorne:1992sdb,Strominger:2014pwa}. See \cite{Strominger:2017zoo} for review of the subject from a modern perspective. As far as we know, the shadow of such black holes has not been intensively studied in the literature.
 
By carefully analyzing the asymptotic fall-off conditions of the metrics of 
the asymptotic flat spacetimes, Bondi, van der Burg, Metzner, and Sachs discovered the underlying asymptotic symmetries, now known as Bondi-Metzner-Sachs (BMS) symmetries \cite{Bondi:1962px,Sachs:1962wk,Sachs:1962zza}. The BMS symmetries are not exact isometries of the spacetime, but are defined by choosing appropriate fall-off conditions for the metric components. From Noether's theorem, the BMS symmetries give rise to conserved charges, which satisfy the so-called BMS algebra. Besides the charges associated with the asymptotic Poincaré symmetry, the so-called hard charges, these conserved charges also include an infinite number of ``soft'' charges, which can be thought of as the localized modes of momentum and angular momentum on the asymptotic celestial sphere, called by the names of supertranslations and superrotations, respectively. Therefore, these soft charges can be understood as the generators of conformal symmetries on the asymptotic celestial sphere \cite{Pasterski:2016qvg}. Moreover, it was understood that the BMS symmetries, as the symmetries near the future and null infinities, act non-trivially on the asymptotic scattering states \cite{Strominger:2013jfa} thus S-matrix  via the associated Ward identities \cite{He:2014laa} to yield the soft theorem discovered long ago by Weinberg \cite{Weinberg:1965nx}, stating the factorization of the soft factor from the S-matrix.

Since there is no local observable in the general covariant theory such as GR, all BMS conserved charges are defined non-locally. Thus, these charges will also dress the other asymptotic boundaries, such as ``hairs" on the horizons of black holes \cite{Pasterski:2020xvn}. Just like the hard charges become the hard hairs of the black holes, the soft charges will then be dressed as the supertranslation/superrotation soft hairs and change the near horizon geometry. The black holes with linear soft hairs are in agreement with the leading terms of the BMS metrics \cite{Hawking:2016msc,Hawking:2016sgy}. However, it is nontrivial to construct black holes with soft hairs as the exact vacuum solutions of the Einstein equations. For 4D GR Compère and Long obtained exact black hole solutions with supertranslation hairs by novel coordinate transformations \cite{Compere:2016hzt}. In \cite{Chakraborty:2017pmn,Raju:2020smc}, fundamental issues related to black hole physics like information paradox and its possible solutions are discussed.

Unlike the black holes, the usual stars have no event horizon to support the soft hairs. This is because that the soft hair generated by the singular coordinate transformation can, however, be moved freely outside the stars to yield no physical effect. Therefore, there is no nontrivial supertranslation state in the usual solar-system like regime to be observed by the light-bending tests.  On the other hand, the supertranslation effect can be encoded in the gravitational memory in the usual gravitational wave observations of the mergers of binary black holes \cite{Thorne:1992sdb,1987Natur.327..123B}. However, such memory effect is of higher post-Newtonian order, and it requires better resolution than the one of the current LIGO/Virgo/KAGRA interferometers to observe it unambiguously.  The other way to observe the supertranslation hair as the near-horizon effect of a black hole is through black hole shadow observations, as we will discuss in this work.

Some properties of the above supertranslated black holes have been studied, e.g., the proposal of utilizing entanglement of the soft modes to resolve the information paradox \cite{Hawking:2016msc,Bousso:2017dny,Bousso:2017rsx,Haco:2018ske}, the Hawking radiation flux \cite{Chu_2018,Compere:2019rof,Lin:2020gva} and the flux-balance laws \cite{Compere:2019gft}. In this work, we will study the black hole shadow of such supertranslated black holes. Naively, one may expect that any observable such as the size or shape of the shadow should be the same as the Schwarzschild counterpart since the supertranslated black holes are just obtained from the Schwarzschild black hole by large gauge transformations. However, due to the soft hairs, the polar symmetry is broken so that the null condition, i.e., the Hamilton-Jacobi equation of the null geodesic looks very complicated. This implies that the effect of the soft hair on the black hole shadow could be nontrivial.  Moreover, the shadow is some kind of non-local observable, and it is not clear how it will be affected by the soft hair generated by the large gauge transformation. This is also an interesting issue we would like to examine in this work.

The rest of the paper is organized as follows. In section \ref{sec:2}, we summarize the properties of supertranslated black holes. 
In section \ref{sec:3}, as a precursor study we consider the analytical construction of the shadows for the black holes with linear supertranslation soft hairs. In section \ref{sec:4}, we adopt the numerical ray-tracing method to construct the shadows for the supertranslated black holes with the full non-linear soft hair and compare their features with the Schwarzschild ones. Finally, we conclude our paper in section \ref{sec:5}. Some technical details are given in the appendices.  In Appendix \ref{append-geos}, we lay down the null geodesic equations in the background metric of the supertranslated black hole, and obtain the separated equations from Hamilton-Jacobi equation. In Appendix \ref{append-a}, we sketch the analytical method of shadow construction adopted in section \ref{sec:3}.  In Appendix \ref{appendix-c} we sketch an accretion model adopted in section \ref{sec:4} to construct the accretion pattern near the supertranslated black hole.

\section{Supertranslated Black Holes}\label{sec:2}

In this paper, we would like to consider the shadow formed by the light geodesics passing by a black hole carrying  soft hairs. There are two kinds of soft hairs, the supertranslation ones, and the superrotation ones \cite{Hawking:2016sgy}. However, the explicit metric has only been constructed for the black hole carrying the supertranslation hairs but not the superrotation ones. For our purpose, we will also focus on the supertranslated black holes in this paper.  

The supertranslated black holes are first proposed and studied by Hawking, Perry, and Strominger in \cite{Hawking:2016msc}. They also showed that the soft hairs can be implanted by shock waves. However, their black hole carries the soft hair only up to linear order. For the first time, Comp\`ere and Long \cite{Compere:2016hzt} constructed the full metric of the supertranslated black hole  by applying a nontrivial coordinate transformation to the metric of Schwarzschild black hole in isotropic coordinate so that the resultant metric still solves the vacuum Einstein equation. This soft hair of the metric in \cite{Compere:2016hzt} is encoded in an arbitrary function on the celestial sphere denoted as $C(\theta,\phi)$. For simplicity, we will assume the $C$-function is $\phi$ independent. In this case, Iofa showed that the metric can be put into a neat Comp\'ere-Long-Iofa (CLI) form \cite{Iofa:2018pnf, Iofa:2019jjm}: 
\be\label{CLI}
ds^2 = -Vdt^2+\frac{\sec^2\psi}{V}dr^2 + 2\frac{(1-\psi')\tan\psi}{\sqrt{V}}dr d\theta + r^2[(1-\psi')^2 d\theta^2 + \sin^2(\psi-\theta)d\phi^2].
\ee
where 
\bea 
V(r) &=& 1-2M/r, \\
\psi(r, \theta) &=& \sin^{-1}[\frac{2C'(\theta)}{K(r)}], \\
K(r) &=& r-M+\sqrt{r(r-2M)}
\eea
with $':= \frac{d}{d\theta}$ and $M$ the mass of the black hole. This CLI metric is reduced to the Schwarzschild black hole metric for $C(\theta)=0$ as expected. If we expand the CLI metric up to the first order in $C$, this metric will be equivalent to the static case of the one in \cite{Hawking:2016msc}.

The Killing horizon of the CLI metric is determined by $g_{tt}=0$, thus is located at $r=2M$ as its Schwarzschild counterpart. However, the metric is no longer spherically symmetric, so is the near horizon geometry, and there exists some new type of coordinate singularity or even the naked singularity. For example, to avoid the appearance of naked singularities, it requires \cite{Iofa:2018pnf,Compere:2016hzt}, 
\be\label{wccc}
|C'|\le \frac{M}{2} \qquad \mbox{or} \qquad |\sin\psi|<1\;.
\ee 
Otherwise, it yields $g_{rr}<0$ for $r>2M$. On the other hand, a new coordinate singularity characterized by the collapse of the spatial 3-metric, i.e., $\det g_{ab}^{(3)}=0$ defines the so-called supertranslation horizon at $r=r_{SH}$. Explicitly, $r_{SH}$ is determined by the following condition 
\be
4\left(C'^2+C''^2\right)= K^2(r_{SH}).  \label{supertranslation_horizon}
\ee 
Note that the supertranslation horizon is nonspherical so that $r_{SH}(\theta)$ is angular dependent. This nonspherical feature due to the soft hairs seems to violate the no-hair theorem. However, the soft hairs will not change the hard hairs of the black hole and are the mere effect of the large gauge transformation. Especially, one can check that the Hawking temperature and the associated Hawking fluxes remain the same as those for the Schwarzschild black hole \cite{Chu_2018,Compere:2019rof, Lin:2020gva}. This then makes our study of shadow interesting, as it is not so clear if the nonlocal nature of the soft hairs can be detected by the black hole shadow, which is kind of semilocal observable. On the other hand, in reality, the black hole will be surrounded by the accreting matter, whose hydrodynamics will then be affected by the near horizon soft hairs, and yield nontrivial shadow. Therefore, the soft hairs will definitely give observable imprints on the shadow of the black hole.

\begin{figure}
\centering
\begin{subfigure}{.5\textwidth}
  \centering
  \includegraphics[width=0.85\linewidth]{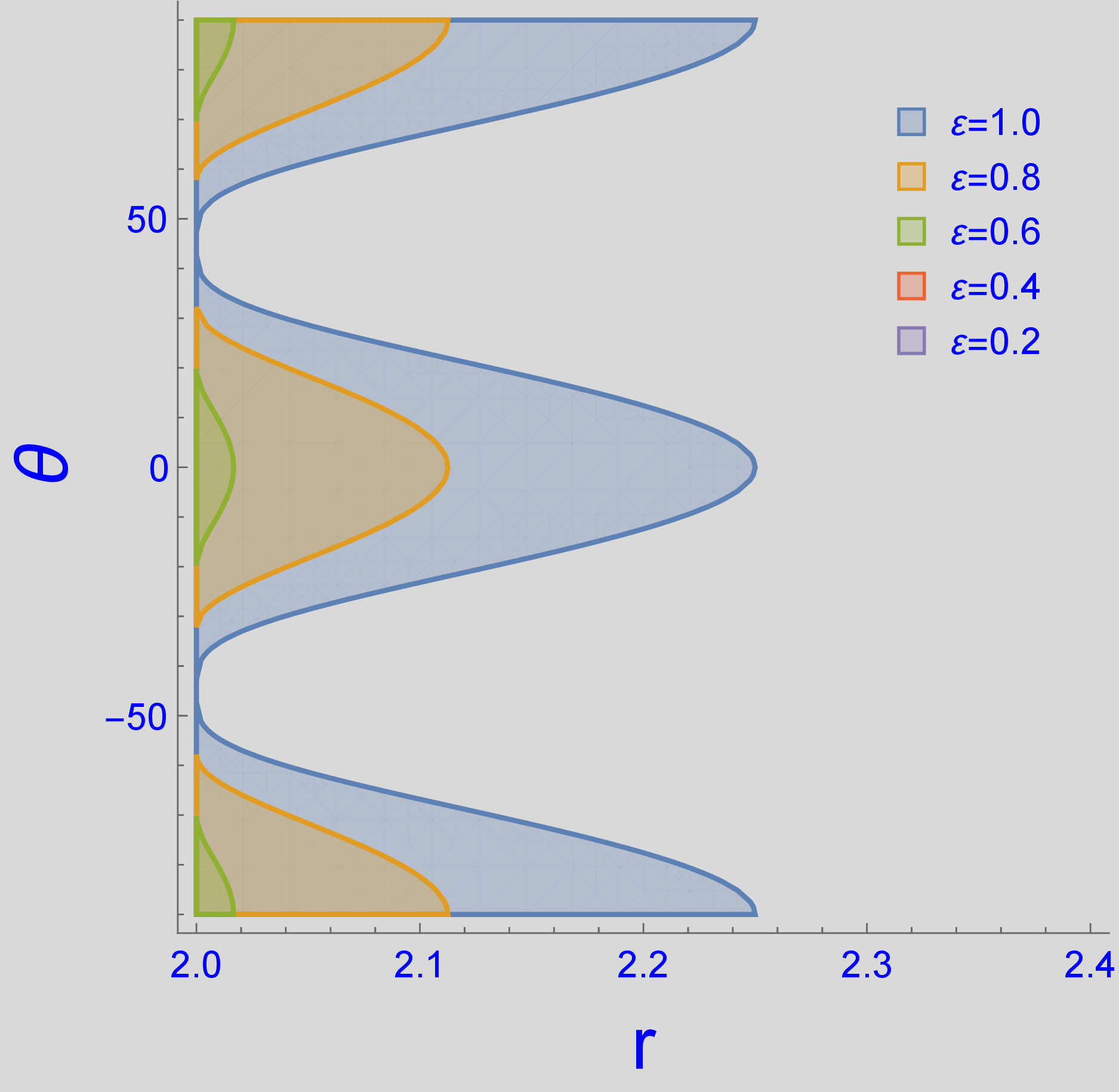}
  \caption{For $C\sim {\mathcal E} Y_{20}$}
  \label{fig:sub1}
\end{subfigure}%
\begin{subfigure}{.5\textwidth}
  \centering
  \includegraphics[width=0.85\linewidth]{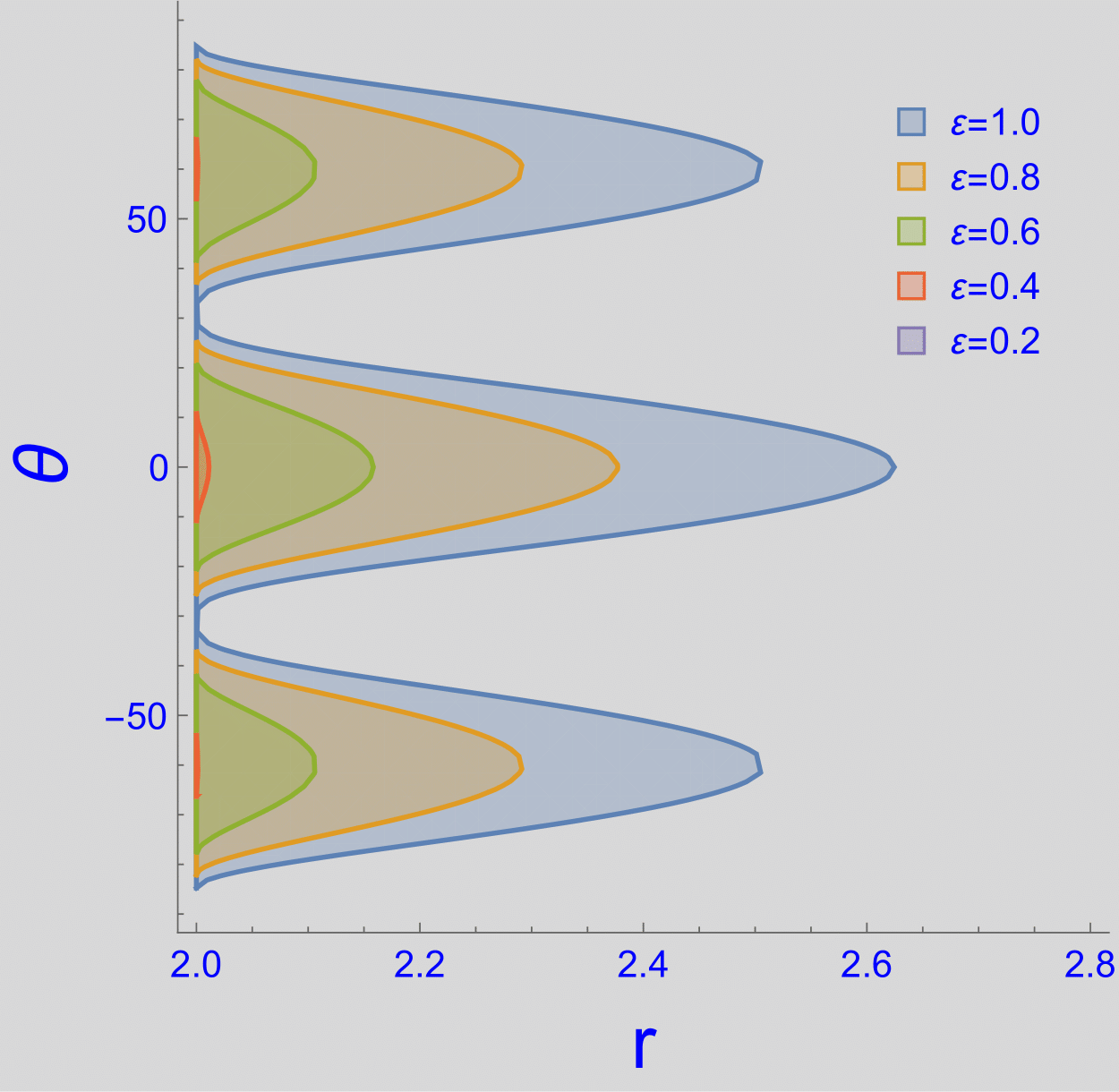}
  \caption{For $C\sim {\mathcal E} Y_{30}$}
  \label{fig:sub2}
\end{subfigure}
\caption{Angular dependence of the supertranslation horizon $r_{SH}(\theta)$ for the choice of $C(\theta)$ given in \eq{C_Yl0} with $\ell=2$ (Left panel) and $\ell=3$ (Right panel). Different colors indicate different values of $\mathcal E$ as shown by the legends. Note that the supertranslation horizon is outside the Killing horizon at $r=2M$ but inside the photon sphere of the Schwarzchild black hole at $r=3M$. On the other hand, for the case with $\ell=1$ (not shown) the supertranslation horizon is inside the Killing horizon and unobserved.}
\label{fig:SH_Y}
\end{figure}

In this work, we will consider the black hole without the naked singularity to avoid the violation of the weak cosmic censorship conjecture \cite{Penrose:1969pc}, so that the soft hair function $C(\theta)$ should satisfy \eq{wccc}.  Since $C$ can be quite generic, for concreteness we choose the typical spherical harmonics $Y_{\ell,m}$, i.e., we will choose the following $C$'s for the consideration of shadow in this work,
\be\label{C_Yl0}
C(\theta)= \mathcal{E} N_{\ell} Y_{\ell,0}(\theta), \quad \mbox{for } \ell=1,2,3,
\ee
where $N_{\ell}$'s are the normalization constants so that the bound \eq{wccc} is saturated for ${\mathcal E}=1$. In Fig. \ref{fig:SH_Y}, we have shown the angular dependence of the position of the supertranslation horizon for the choice of $C(\theta)$ given in \eq{C_Yl0} for $\ell=2,3$. We see that the supertranslation horizons for these two cases are outside the Killing horizon at $r=2M$, but still inside the  photon sphere of the Schwarzschild black hole at $r=3M$. We did not show the case of $\ell=1$ because its supertranslation horizon is below the Killing horizon and unobservable. This implies that the supertranslation horizon may not directly change the shape of the shadow. On the other hand, the supertranslation horizon can exceed the photon sphere at $r=3M$ if $\mathcal E>1$ to directly change the shape of the black hole shadow. However, in such cases, there appear naked singularities and the CLI metric no longer describes a black hole.

In the following, we will discuss the light-ray behavior around this CLI metric of the supertranslated black hole to examine the effect of soft hairs on the black hole shadow. As a precursor study, we first consider the CLI metric up to the linear order of soft hair function $C$ and construct the black hole shadow by the analytic method. Finally, we construct the black hole shadow for the full CLI metric by the numerical ray-tracing method.

\section{Analytical study of black hole shadow with linear soft hair}\label{sec:3}

To set up the framework of the black hole shadow construction as a precursor study, we first consider the case with linear supertranslation hair. We like to examine if the shadow problem can be analytically solved by proceeding with the linear perturbation around the known shadow of the Schwarzschild black hole. Although the CLI metric is diffeomorphic to the Schwarzschild metric, it breaks the polar symmetry even at the linear order of $C$ so that the integrability or separability of the null geodesic equations seems lost. However, the integrability could be hidden. Thus, we can obtain the CLI counterpart of the separable equations of the Schwarzschild metric by the coordinate transformation, although these separated equations are not decoupled. In this section and in the Appendix \ref{append-geos} and \ref{append-a}, we explicitly obtain the separated geodesic equations and show that the black hole shadow can be constructed analytically with the help of the above separated geodesic equations. From the results, we will see that the black hole shadow will be affected by the linear soft hair. The effect of the full nonlinear soft hair on the black hole shadow will be studied by the numerical ray-tracing method in the next section.

\begin{figure}[!ht]
  \centering
 \includegraphics[scale=0.18]{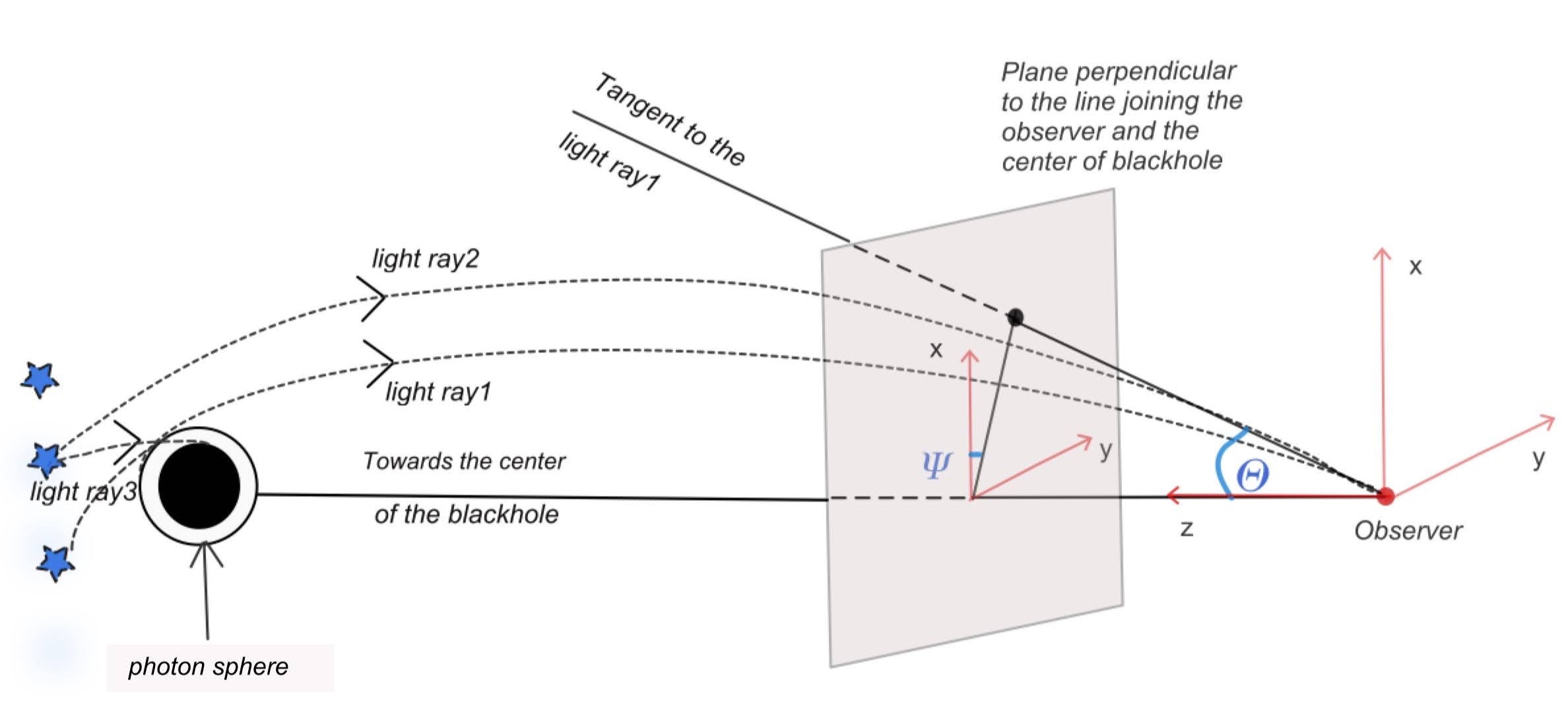}
\caption{Schematic picture of black hole shadow on the celestial image plane left by the light-rays marginally escaping from the black hole and ending on the observer. Light ray 2 slightly bends due to gravitational pull of the black hole and flies to the infinity. Light ray 3 falls into the black hole and never reaches the observer. Lastly, light ray 1 which corresponds to the boundary of the shadow flies to the infinity and tangential to the photon sphere at some point in their trajectories.}
\label{fig:schematic} 
\end{figure}

Due to the strong gravity pull, a light-ray can be captured by the black hole if the impact parameter is not large enough. The shadow cast by the black hole on the celestial plane/sphere is schematically shown in Fig. \ref{fig:schematic}. 
 The procedure to construct the shadow is to trace the trajectories of the light-rays which are tangential to the photon sphere to the observer, and then mark down the positions of these light-rays when crossing the celestial image plane or sphere.  When the geodesic equations are integrable, each light ray is uniquely determined by the conserved quantities, i.e., energy and angular momentum. Therefore, those light-rays coming from just outside the photon region will be characterized by specific energy and angular momentum. We can then match the tangent to this light-ray expressed in black hole coordinate representation with the same written in terms of local tetrads at the position of the observer to analytically determine the position of the shadow in the observer's sky. This is the case for the Schwarzschild and Kerr black holes \cite{Grenzebach:2014fha}. For the supertranslated black hole described by the CLI metric, the polar symmetry is broken so that the geodesic equations are not integrable. Despite that, in the following, we will solve the photon sphere up to the linear order of supertranslation hair and check if the shadow can be constructed analytically.

The light-ray geodesic equation can be obtained from the variation of the following Lagrangian:
\be
{\cal L}=\frac{1}{2} g_{\mu\nu} \dot{x}^{\mu} \dot{x}^{\nu}
\ee
where $\dot{} :=\frac{d}{d \tau}$ with $\tau$ the worldline time. Because the light-ray is null, so we should impose the null condition, which up to the first order of $C$ takes the form (Note in this section we will set $M=1$ for simplicity.):
\bea
 0 &=&  \left(L_z^2 \csc^2\theta - \frac{E^2 r^3}{r-2}\right)+ \frac{r^3 \dot{r}^2}{r-2} + r^4 \dot{\theta}^2 \nn \\ 
&& + \frac{4}{(r-1+\sqrt{r(r-2)})}\Bigg(\Big( L_z^2 \cot\theta \csc^2\theta + \sqrt{\frac{1}{r(r-2)}} r^4 \dot{r} \dot{\theta} \; \Big) C'-r^4 \dot{\theta}^2 C''\Bigg)\;. \label{HJeq_C}
\eea 
Since the metric is static and independent of $\phi$, the energy $E$ and $z$-component of the angular momentum $L_z$ are conserved and are defined by 
\bea
E&:=& -\frac{\delta {\mathcal L}}{\delta \dot{t}}=\left(1-\frac{2}{r}\right) \dot{t}\;,\\
L_z &:=& \frac{\delta {\mathcal L}}{\delta \dot{\phi}} =\Big(1- \frac{4C' \cot{\theta}}{K} \Big)  r^2 \sin^2\theta \dot{\phi} \;. 
\eea
Furthermore, by varying ${\cal L}$ we can further obtain the geodesic equations. However, due to the complication of the CLI metric, the geodesic equations are quite involved and not very illuminating. In Appendix \ref{append-geos} we have written down the geodesic equations for the full CLI metric and explained the procedure of solving them for the null geodesics with given initial conditions. Besides, for this section we also expand the geodesic equations up to the linear order in $C$, which one can find in \eq{t-geo-C}-\eq{phi-geo-C} of Appendix \ref{append-geos}.

Using the geodesic equations \eq{t-geo-C}-\eq{phi-geo-C} and null condition \eq{HJeq_C}, we can solve the conditions for the photon sphere, i.e., $\dot{r}=\ddot{r}=0$. The results are as follows:
\be 
r_p=3, \qquad \theta_p=\theta_0 + \theta_1 + {\cal O}(C^2)\label{geo_th1}
\ee
where the $\theta_0$ and $\theta_1$ are the leading and first order solutions of the following two equations respectively,
\bea
\dot{\theta}_0 &=& \pm \frac{1}{9}\sqrt{27 E^2 -L_z^2 \csc^2\theta}_0\;, \\
\dot{\theta_1}&=& \pm \frac{L_z^2 \cot\theta_0 \csc^2\theta_0 \Big(\theta_1-(4-2\sqrt{3})C' \Big)}{9 \sqrt{27E^2 -L_z^2 \csc^2\theta_0}}+\frac{4-2\sqrt{3}}{9} \sqrt{27E^2 -L_z^2 \csc^2\theta_0}\;.
\label{geo_th2}
\eea 
The leading order result is the same as the ones for Schwarzschild black hole, it is interesting to see that the radius of the photon sphere remains the same at all orders.

With the photon sphere determined, the remaining work to construct the shadow is to find the light-rays connecting the observer and tangential to the photon sphere. Following the method of \cite{Grenzebach:2014fha,Perlick:2021aok} to analytically construct the black hole shadow, we need two linearly independent equations to determine the position of a light-ray on the celestial image plane. The detail of such construction is sketched in Appendix \ref{append-a}. It turns out that it is necessary to separate the null condition, as in the cases of Schwarzschild or Kerr black holes, to obtain these two linearly independent equations. The separability relies on the integrability of the geodesic dynamics, which needs the additional conserved quantities such as the Carter constant. Even though the CLI metric seems breaking the polar symmetry, it is obtained from the Schwarzschild metric by the large diffeomorphsim so that the scalar quantities such as the Carter constant should be preserved. In Appendix \ref{app_Carter} we obtain the counterpart of two separated equations from the null condition of the Schwarzschild metric by performing the corresponding large diffeomorphsim. Although these two resultant equations are non-decoupled, they are helpful for obtaining the linearly independent equations in determining the position of a light-ray on the image plane.

\begin{figure}
\centering
\begin{subfigure}{.5\textwidth}
  \centering
  \includegraphics[width=0.85\linewidth]{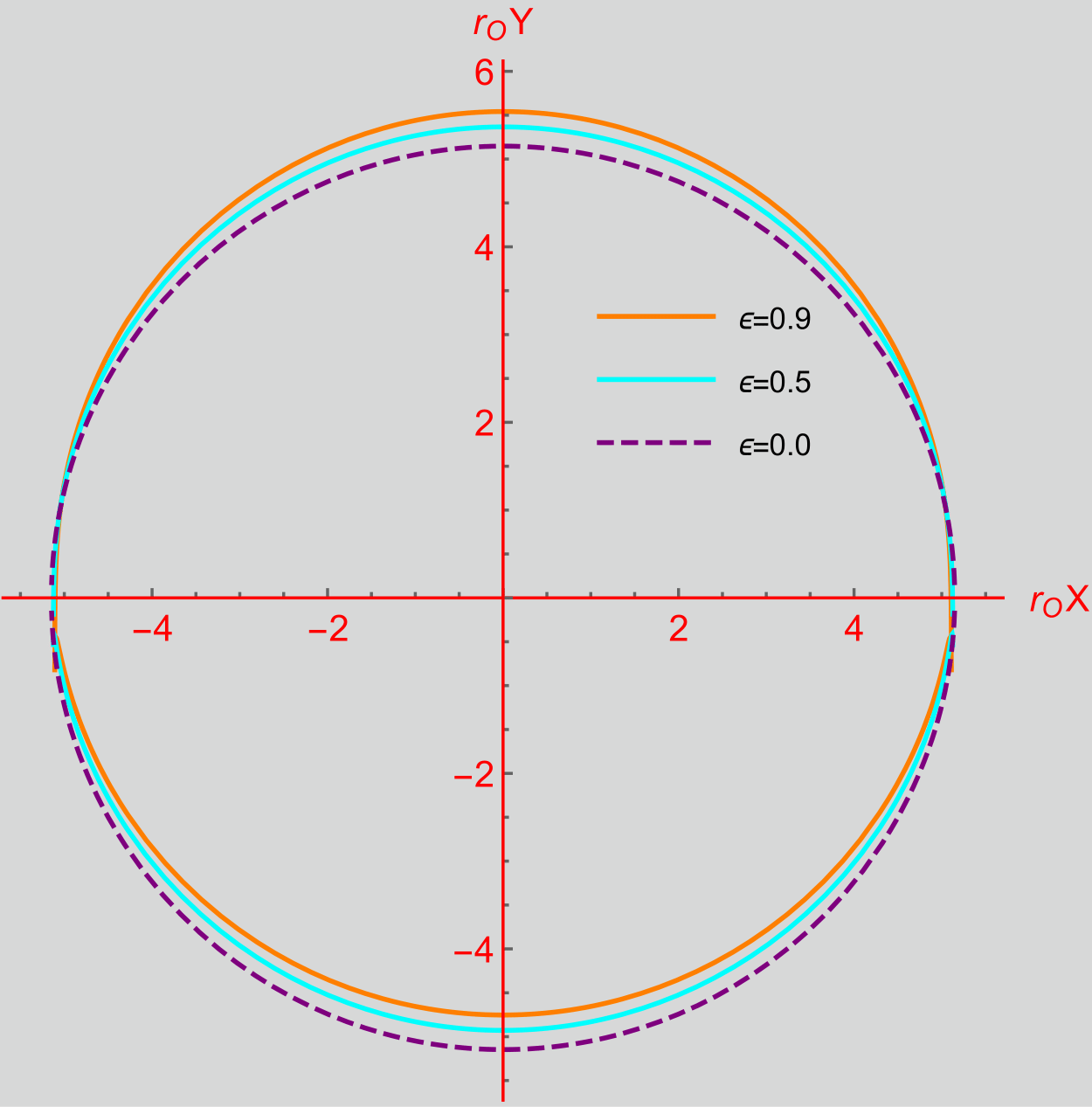}
  \caption{For $\theta_{O}=30^{\circ}$}
  \label{fig:ASsub1}
\end{subfigure}%
\begin{subfigure}{.5\textwidth}
  \centering
  \includegraphics[width=0.85\linewidth]{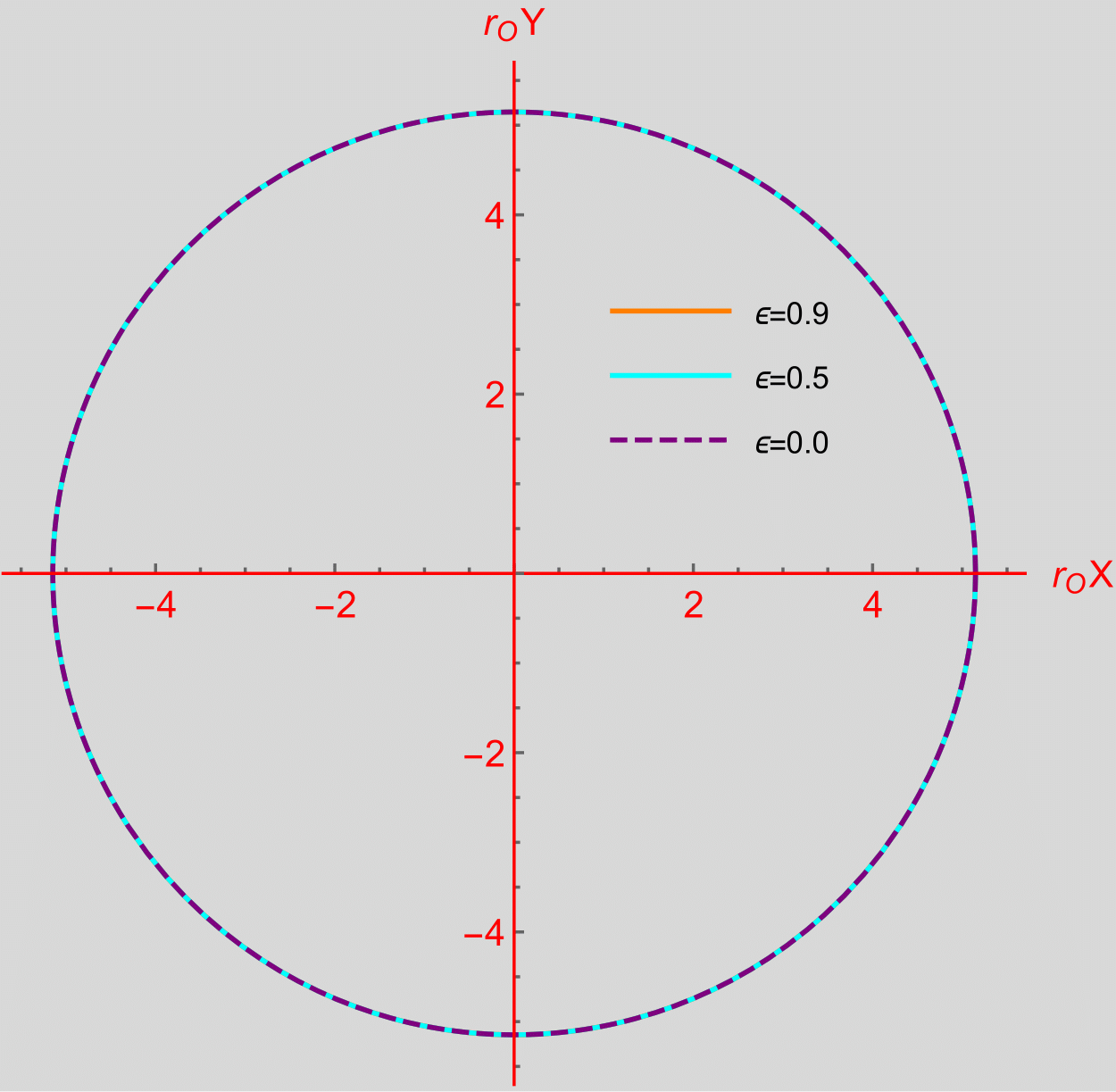}
  \caption{For $\theta_{O}=90^{\circ}$}
  \label{fig:ASsub2}
\end{subfigure}
\caption{Shadow of a black hole with soft hair $C\sim {\mathcal E} Y_{20}$. The observer is situated at $r_O=100 M$. This analytical shadow is parametrized by $X$ and $Y$ defined by \eq{sdmap}. The method for analytical construction of the shadow is also described in Appendix \ref{append-a}.}
\label{fig:analyticalshadow}
\end{figure}

Based on the above result, we can obtain the black hole shadow for the non-rotating black hole with linear supertranslation hair analytically without implementing the numerical ray-tracing method. The above construction and result are given in Appendix \ref{append-a}, and a typical plot of black hole shadow is shown in Fig. \ref{fig:analyticalshadow} for the choice of $C(\theta)$ given in \eq{CII} for observer situated at $r_O=100M$ and $\theta_{O}=30^{\circ}$ and $90^{\circ}$.
In both the figures \ref{fig:ASsub1} and \ref{fig:ASsub2}, the magenta dashed curve shows the boundary of the shadow of a Schwarzschild blac khole. In Fig.~\ref{fig:ASsub1}, we can see that the shadow of a supertranslated black hole is shifted vertically. In the next section, we will see that it is consistent with the shadow obtained by ray-tracing method. From the observational point of view, the shape deformation is the important factor to study. In this section and the next section, we calculate shape asymmetry $\mathcal{A}$ as defined in \eq{def_features2}. The shape asymmetry for Orange and Cyan curve in Fig.~\ref{fig:ASsub1} is $\mathcal{A}=0.0188$ and $0.0105$ respectively. On the other hand, in Fig. \ref{fig:ASsub2}  we see that the shadows remain intact by the linear soft hair for $\theta_{O}=90^{\circ}$. In the next section, we will see that the shadows of the full CLI metric bear the similar features shown above for the linear CLI case.

\section{Shadow of supertranslated black hole}\label{sec:4}

\begin{figure}[!ht]
  \centering
 \includegraphics[scale=0.36]{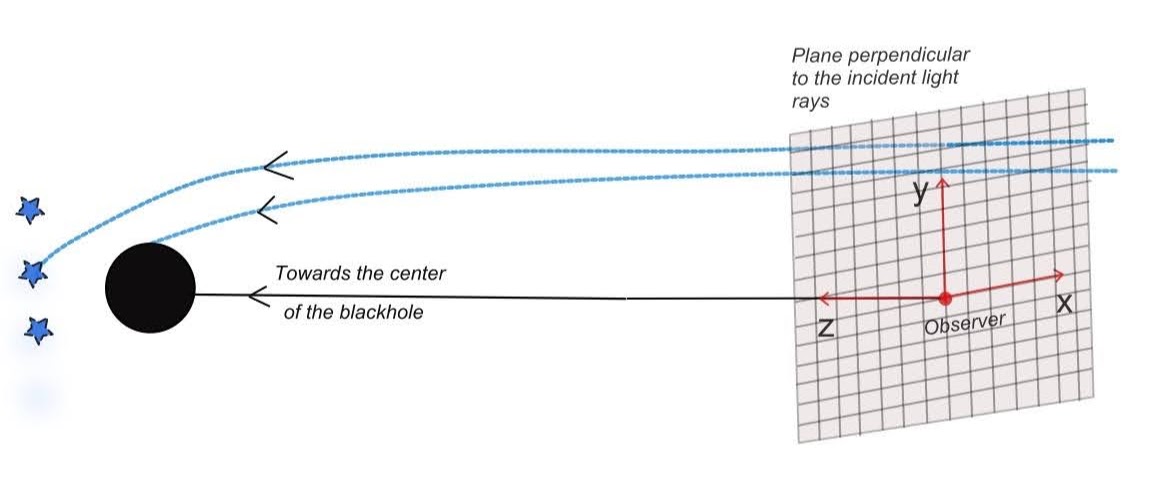}
\caption{Construction of the  shadow of a  black hole for a distant observer by tracing the light-rays, which arrive each pixel on the observer's image plane, backward in time. The $(x,y,z)$ is the observer's local coordinate frame whose origin is given by $(r_O, \theta_O, \phi_O)$ in the black hole's reference frame. The pixel's position w.r.t. black hole frame in terms of its position w.r.t local $(x,y,z)$ frame is given in \eq{r0-eq}.}
\label{fig:raytracing} 
\end{figure}

In this section, we numerically calculate the shadow of the black hole with soft hair, which is described by the full CLI metric \eq{CLI} for the choices of the soft hair functions specified in \eq{C_Yl0}. After specifying the initial condition for rays at the observer's image plane on the celestial sphere, we compute the null geodesics to trace each ray backward in time. The above ray-tracing procedure is schematically illustrated in Fig. \ref{fig:raytracing}.


\subsection{Ray-tracing and Numerical Setup}

To delineate the shadow, we first need to choose a local coordinate system around the observer to label the grid on the celestial image plane. We denote the observer's coordinate axes as $(x,y,z)$ with $z$-axis oriented towards the center of the black hole, of which the local coordinate is denoted by $(r,\theta,\phi)$ used for the CLI metric \eq{CLI}. The observer is located at the origin in her/his local frame, and at $(r_O,\theta_O,\phi_O)$ in the black hole's frame. 

Unlike the consideration in section \ref{sec:3} and Appendix \ref{append-a} for which the black hole can be at any distance from the observer, here we are only interested in the shadow of a distant black hole. The shadow will then be located only in a small region on the celestial sphere, which can be approximated by a celestial image plane perpendicular to the $z$-axis. Moreover, the curvature effect due to the distant black hole is negligible so that the light-rays will end almost perpendicular to the image plane, i.e., 
\be\label{init_vec}
(\dot{x},\dot{y}, \dot{z})\approx (0,0,1)\;.
\ee 
The above fact will not change even after including soft hair because the size of the soft hair is bounded by black hole mass, i.e., \eq{wccc}, to avoid naked singularities. 

For simplicity, we can assume the image plane is located at $z=0$ and is described by the grids labeled by $(x,y)$. For notational consistency with other literature, we denote $(x,y)$ plane by $(\alpha,\beta)$ plane in the latter calculation of shadow. Along with \eq{init_vec} it gives the initial conditions for the ray-tracing. However, as the geodesic equations for the light-rays are written in the $(r,\theta,\phi)$ coordinates of the CLI metric, we need to translate the initial conditions of the light-ray on the image plane in the observer's frame to the ones in the black hole's frame. Following \cite{younsi14, younsi16,Pu:2016eml}, the label $(x,y)$ of the grids on the image plane corresponds to the initial position in the black hole frame as follows:  
\bea
r(0)&=&\sqrt{r_O^2+x^2+y^2}\;, \label{r0-eq} \\
\theta(0) &=& \cos^{-1}\Big[\frac{r_O \cos\theta_O+y \sin\theta_O}{r(0)}\Big]\;,\\
\phi(0) &=& \tan^{-1}\Big[\frac{(y\cos\theta_O-r_O\sin\theta_O)\sin\phi_O - x \cos\phi_O}{(y\cos\theta_O-r_O\sin\theta_O )\cos\phi_O + x \sin\phi_O}\Big]\;,
\eea
and the initial tangent vector \eq{init_vec} is translated into
\bea
\dot{r}(0)&=&-\frac{r_O}{r(0)}\;, \\
\dot{\theta}(0)&=&\frac{(x^2+y^2) \cos\theta_O - y r_O \sin\theta_O}{r^2(0)\sqrt{x^2+(y\cos\theta_O-r_O\sin\theta_O)^2} }\;, \\
\dot{\phi}(0)&=&\frac{x\sin\theta_O}{x^2+ (y\cos\theta_O-r_O\sin\theta_O)^2}\;.
\label{phid0-eq}
\eea

The null geodesics \eq{r-geo}-\eq{phi-geo} are numerically computed by solving the set of eight equations\footnote{the eight equations are associated with ($\dot{t}$,  $\dot{r}$, $\dot{\theta}$, $\dot{\phi}$; $\ddot{t}$,  $\ddot{r}$, $\ddot{\theta}$, $\ddot{\phi}$), or, equivalently, ($p^{t}$,  $p^{r}$, $p^{\theta}$, $p^{\phi}$; $\dot{p}^t$,  $\dot{p}^r$, $\dot{p}^{\theta}$, $\dot{p}^{\phi}$).} which consists of ($p^{\gamma}=\dfrac{dx^{\gamma}}{d\lambda}, \dfrac{dp^{\gamma}}{d\lambda}$), where $\gamma=(t,r,\theta,\phi)$ and $\lambda$ denotes the affine parameter. The massive computations are performed by a GPU-based ray-tracing code\footnote{\href{https://github.com/hungyipu/Odyssey}{https://github.com/hungyipu/Odyssey}} {\it Odyssey} with the Runge-Kutta method \cite{Pu:2016eml}. Along each ray, there are two conserved quantities: the angular momentum $L_z\equiv p_{\phi}$ and the energy $E\equiv -p_{t}$. These two conserved quantities can be specified by applying the initial condition to \eq{E-eq} and \eq{L-eq}. As mentioned, we do not supply this conserved quantities for solving the geodesics. Instead, we monitor the difference between the conserved quantities and the numerically computed $p_{\phi}$ and $p_{t}$ to make sure the numerical error along the geodesic computations are under control.

Starting from a specific pixel on the celestial image plane ($\alpha, \beta$), we trace the ray backward in time until it hits the event horizon ($r=2 M$), or hits the supertranslation horizon ($r\ge 2 M$) (see also Fig. \ref{fig:SH_Y}), or escapes outside region of interest far away form the black hole. The shadow image on the ($\alpha,\beta$) plane can therefore be obtained by collecting the rays which either hit the event or supertranslation horizons.

\begin{figure}[!ht]
  \centering
 \includegraphics[scale=0.4]{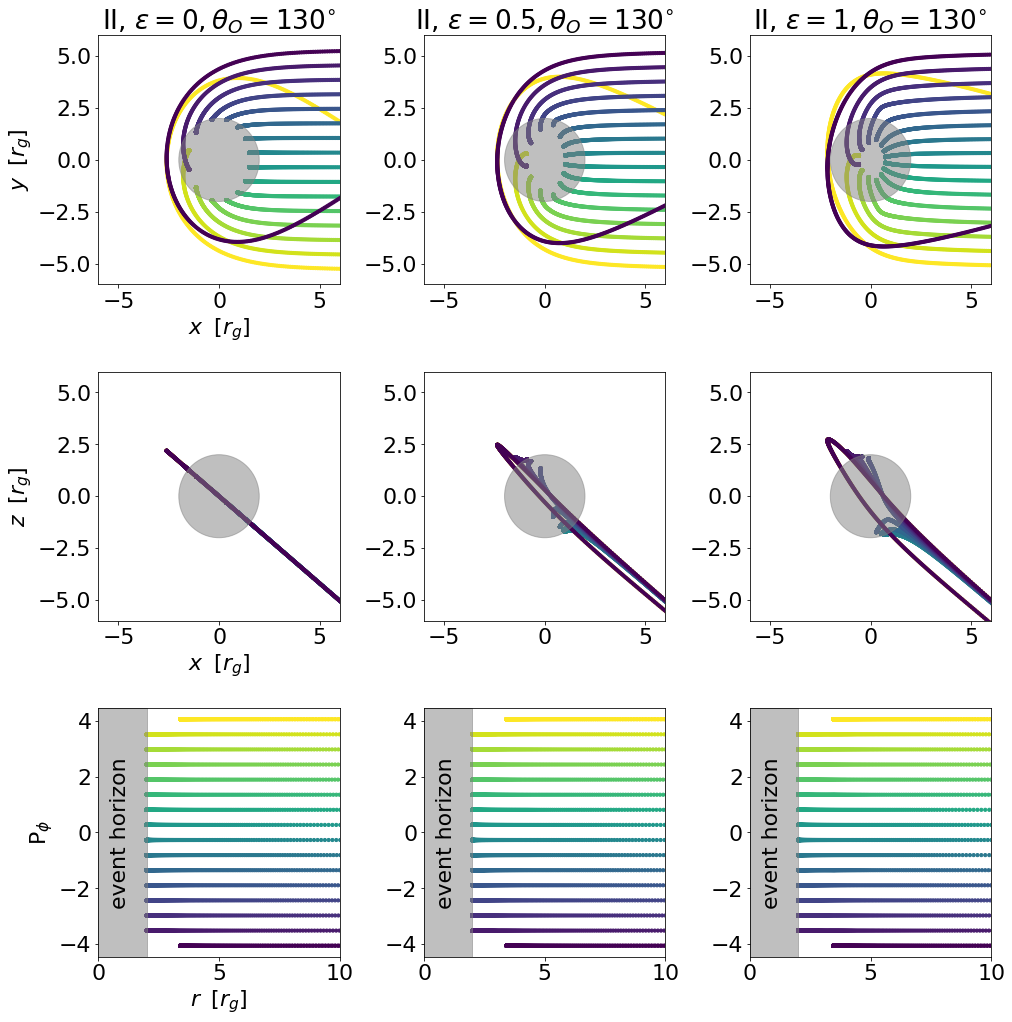}
\caption{Some examples of Light-rays and their properties in the CLI metric with soft hair II given in \eq{CII} with ${\mathcal E}=0$ (first column, Schwarzschild metric), $0.5$ (second column) and $1.0$ (third column). The first row shows the face-on view  of the geodesics. The second row shows the edge-on view and therefore highlights the features caused by non-vanishing $g_{r\theta}$. The third row shows the constancy of angular momentum $L_z$ along the light-rays as a numerical consistency check. The shaded regions indicate the black hole's event horizon.}
\label{fig:geod} 
\end{figure}

In our computation, the field of view of the image is set to be $12M$ by $12M$, with resolution 512 by 512 pixels{\footnote{The choice of such resolution is a compromise between the computation cost and the purpose of exploring the shadow features.}. As mentioned, we calculate the black hole shadow for the soft hair functions given in \eq{C_Yl0}. To be concrete, we write down their explicit forms here:
\bea
\mbox{I:}&& {\mathcal E} M \sqrt{\frac{\pi}{3}} Y_{10}(\theta)=\frac{1}{2} {\mathcal E} M \cos\theta \;, \label{CY10} \\
\mbox{II:}&& {\mathcal E} M \sqrt{\frac{4 \pi}{45}} Y_{20}(\theta)=\frac{1}{6} {\mathcal E} M (3 \cos^2\theta -1)\;, \label{CII} \\
\mbox{III:}&& {\mathcal E} M \sqrt{\frac{15 \pi}{448}} Y_{30}(\theta)=\frac{\sqrt{15}}{32} {\mathcal E} M (5 \cos^3\theta -3\cos\theta)\;. \label{CY30}
\eea 
In the following numerical calculations, we simply set $M=1$.

Compared with the Schwarzschild cases, the CLI metric may result in ``additional motion" of the light ray geodesic along the $\theta$-direction, which is closely related to its nonvanishing $g_{r\theta}$ term (see also Eq. \eq{geo_th1}-\eq{geo_th2}). Such an effect becomes more obvious as the size of the soft hair function is large, i.e., larger $\mathcal E$. The above features are all shown in Fig. \ref{fig:geod} for the soft hair function II. In the next subsection, we will present our numerical results of black hole shadows for the above soft hair functions labeled by I, II, and III with $0\le {\mathcal E} \le 1$ when viewing with different inclination angle $\theta_{O} \in [0^{\circ}, 180^{\circ}]$.

\subsection{Soft-Hair Features of Shadows}

\begin{figure}[!ht]
  \centering
 \includegraphics[scale=0.3]{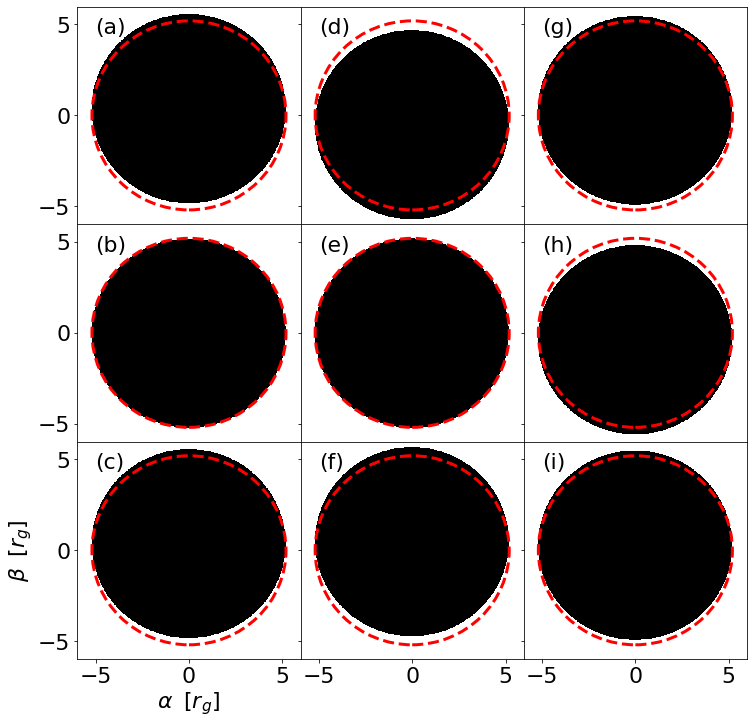}
\caption{Example black hole shadows with soft hairs. The dashed (red) circle is the boundary of the shadow of the Schwarzschild black hole, of which the radius is $\sqrt{27}$. The center of the shadow can be shifted in vertical direction. The same inclination angle (but for different perturbation mode and amplitude) is applied for each row, and the corresponding parameters for these shadows can be found in Fig. \ref{fig:d_beta} according to the labels (a)-(i). Case (b) corresponds to the Schwarchzschild case ($\mathcal{E}=0$).  Note that the shadow for case (e) accidentally coincides with that of Schwarzschild case even if $\mathcal{E}=1$. 
}
\label{fig:shadow} 
\end{figure} 

Some typical examples of shadows of supertranslated black holes are shown in Fig. \ref{fig:shadow}. Due to the limitation on the size of the soft hair to avoid violation of WCCC, from Fig. \ref{fig:shadow} we see that the shadows with soft hairs do not differ much from the Schwarzschild ones (the dashed red circle).

To distinguish the subtle differences and extract the soft-hair features of the shadows, we will examine in detail the following properties of the shadow images: the position shifts $D_{\alpha,\beta}$, the average radius $\bar{R}$, and the shape asymmetry $\mathcal A$, which are defined as follows:
\bea 
D_{\alpha}&\equiv& \dfrac{\alpha_{\rm max}
    +\alpha_{\rm min}}{2}\;, \qquad D_{\beta}\equiv\dfrac{\beta_{\rm max}+\beta_{\rm min}}{2}\;, \\
 \bar{R}&\equiv& \dfrac{\sum\limits_{i=1}^{N} R_{i}}{N}\;,  \qquad 
\mathcal{A} \equiv \sqrt{\dfrac{\sum\limits_{i=1}^{N} (R_{i}-\bar{R})^{2}}{N}}\label{def_features2}\;.
\eea
Here $\alpha_{\rm max/\rm min}$ denotes the rightmost and leftmost horizontal boundary points of the shadow image, respectively. Similarly,  $\beta_{\rm max/\rm min}$ denotes the uppermost and lowermost vertical boundary points, respectively. It is obvious that $D_{\alpha}=D_{\beta}=0$ for Schwarzschild black hole. To calculate $\bar{R}$ and $\mathcal A$, we uniformly pick up $N$ boundary points with $R_i$ ($i=1,2,\cdots, N$ for large enough $N$) their distance from the computed shadow center,$(D_{\alpha},D_{\beta})$. 
For Schwarzschild black hole, $\bar{R}=R_i=\sqrt{27}$ and ${\mathcal A}=0$.

 
Within the numerical error-bar due to the chosen resolution, i.e., $12/512/2\sim0.012M$ in our numerical computation, we find that there is no obvious horizontal shift, i.e., $D_{\alpha}=0$. This is consistent with the axisymmetry of the CLI metric. Moreover, there seems also no shape asymmetry, i.e., $\mathcal{A}=0$ as shown in Fig. \ref{fig:asym} for three types of soft-hair functions given in \eq{CY10}-\eq{CY30} with the size parameter $\mathcal E$ tuned between $0$ and $1$. $\mathcal{A}$ remains under the error bar for most parameter values. This is in accordance with the shadow images shown in Fig. \ref{fig:shadow}, and could be thought of as the observational manifest of the no-hair theorem even in the presence of the soft hairs. 

  
\begin{figure}[!ht]
  \centering
  \includegraphics[scale=0.4]{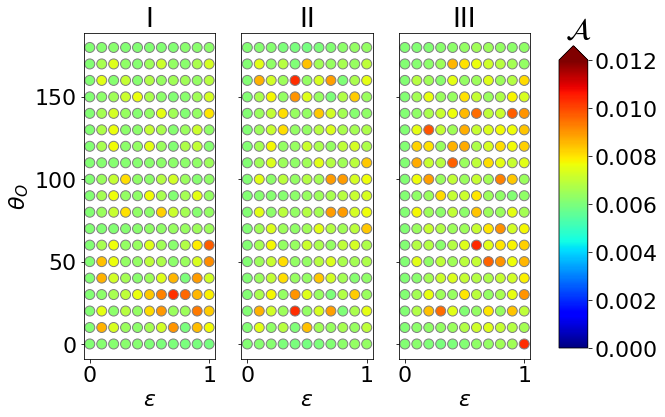}
\caption{Shape asymmetry $\mathcal{A}$ (indicated by the color bar) of the shadows for the three types of soft hair functions given in \eq{CY10}-\eq{CY30} with the size parameter $\mathcal E$ tuned between $0$ and $1$, and for different observer's inclination angles. Within the  error  introduced by the finite resolution about $0.012M$, there is no obvious shape asymmetry. This is some kind of observational manifest of no-hair theorem even in the presence of soft hairs. 
} 
\label{fig:asym} 
\end{figure}

On the other hand, there are obvious vertical shifts $D_{\beta}$ of the shadows, as can be easily seen in Fig. \ref{fig:shadow}. In Fig. \ref{fig:d_beta}, we show $D_{\beta}$ for the soft hair functions given in \eq{CY10}-\eq{CY30} and for different observer's inclination angle, and the corresponding shadow images for the labeled points (a)-(i) are already shown in Fig. \ref{fig:shadow}. The vertical shifts of the shadows are caused by the nonvanishing $g_{r\theta}$ so that the geodesics are no longer planar as pointed out in Fig. \ref{fig:geod}. It is interesting to see that the vertical shifts can be either upward ($D_{\beta}>0$) or downward ($D_{\beta}<0$), depending on the observer's inclination angle and the choice of the soft hair function. From Fig. \ref{fig:d_beta}, it is apparent that $D_{\beta}$ always remains zero for $\epsilon=0$ for all inclination angle because of the spherical symmetry of the Schwarzschild black hole. In the presence of soft hair, the spherical symmetry is broken and therefore $D_{\beta}$ varies with the observer's inclination angle. However, the change in inclination angle only affects the vertical shift, keeping the overall shape of the shadow still circular. In practical, this distinct feature can be observed with multiple sources after taking care of degeneracy due to other source parameters. While the shift of the shadow center depends on the observer's inclination angle, for a specific perturbation mode, there could  be "nodal points" between the red and blue region shown in Fig. \ref{fig:d_beta}, resulting in a net effect which leaves the shadow unchanged. One example is the case (e) shown in Fig. \ref{fig:d_beta} and Fig. \ref{fig:shadow}. To highlight how shadow center could be unaffected with soft-hair perturbation, in Fig. \ref{fig:geod_b} we plot the geodesics for different observer's inclination angle with increasing $\mathcal{E}$, for the perturbation model II.


\begin{figure}[!ht]
  \centering
\includegraphics[scale=0.4]{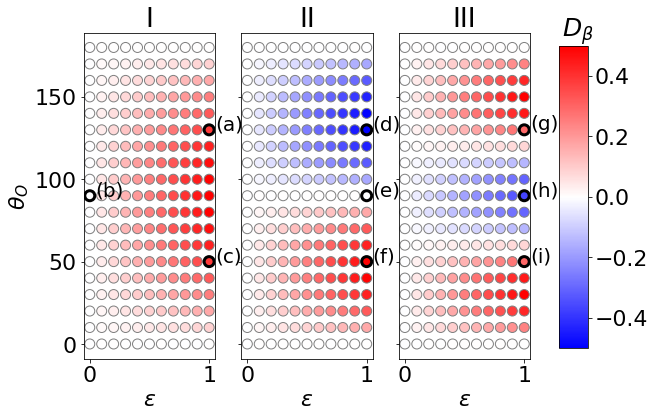}
\caption{Vertical shifts $D_{\beta}$ (indicated by the color bar) of the center of the shadow images for the three types of soft hair functions given in \eq{CY10}-\eq{CY30} with the size parameter $\mathcal E$ tuned between $0$ and $1$, and for different observer's inclination angles.
The shadow morphology for the parameters indicated by (a)-(i) are shown in Fig. \ref{fig:shadow}. } 
\label{fig:d_beta} 
\end{figure}

\begin{figure}[!ht]
  \centering
\includegraphics[scale=0.4]{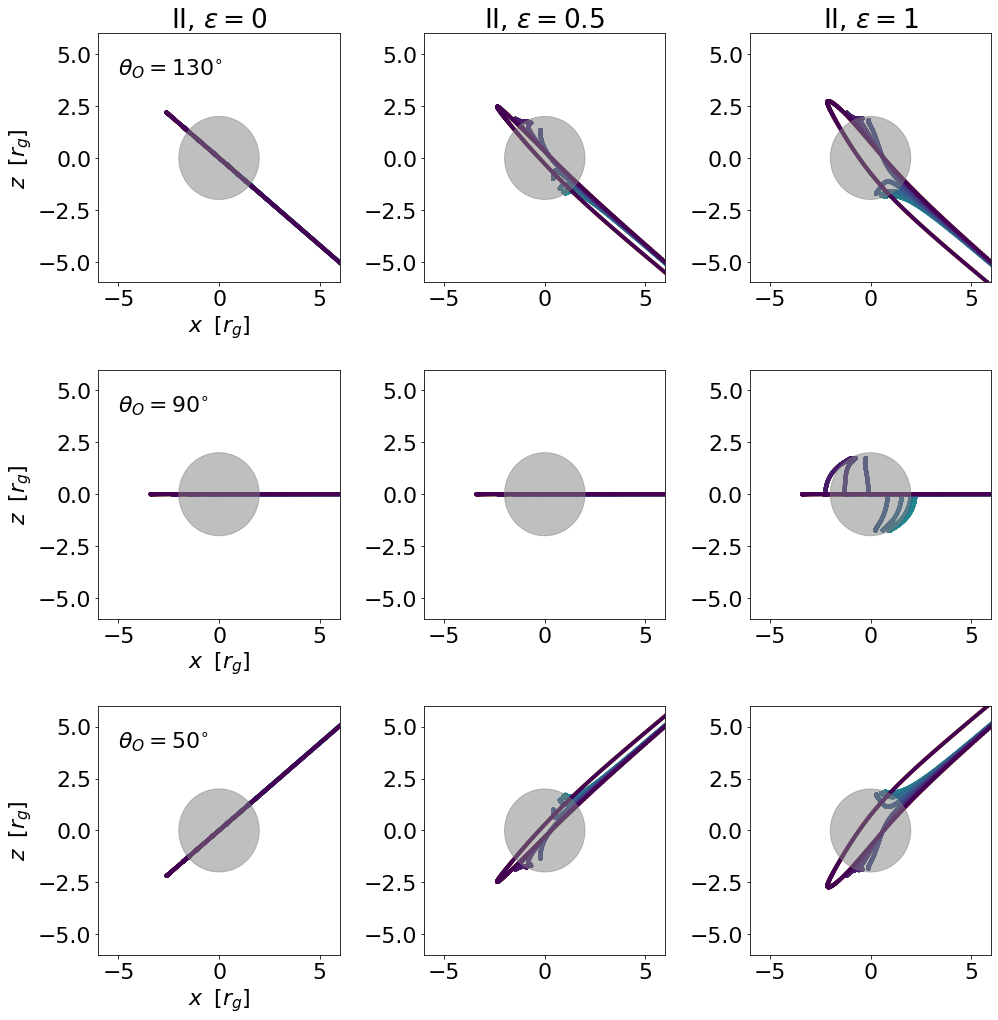}
\caption{Edge-on view of geodesics for different observer's inclination angle and different size parameter $\mathcal{E}$ for type II perturbation. When $\mathcal{E}=0$ (Schwarzschild case), the geodesics are coplanar. As $\mathcal{E}$ increases, the geodesics is no longer coplanar due to the nonzero $g_{r\theta}$ term, resulting in the shift of center of the shadow. The geodesics shown in this figure contribute to the computed shadow image along the $\beta=0$ plane. The right column ($\mathcal{E}=1$) corresponds to cases (d), (e), (f) in Fig. \ref{fig:d_beta}, respectively. The case ($\theta_{O}=130^{\circ}$) shown in Fig. \ref{fig:geod} is identical to the case shown in the first row.} 
\label{fig:geod_b} 
\end{figure}

\begin{figure}[!ht]
  \centering
  \includegraphics[scale=0.4]{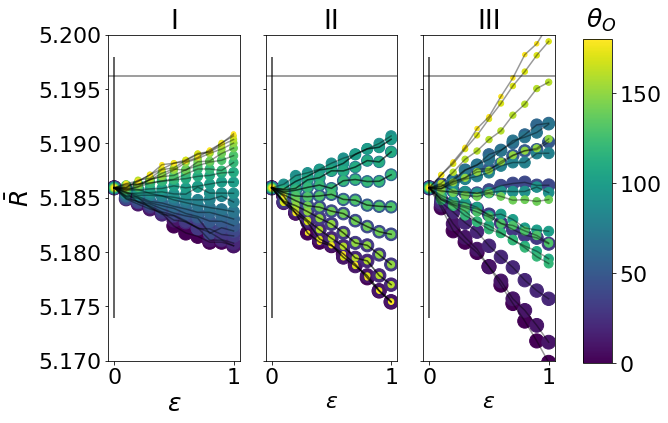}
\caption{Average radii $\bar{R}$ of the shadows for three types of soft hair functions given in \eq{CY10}-\eq{CY30} with $0\le {\mathcal E} \le 1$. The long vertical line indicates the numerical error bar $\pm 0.012$, within which our numerical value $\bar{R}_0\simeq 5.186$ for the Schwarzschild case is consistent with the theoretical one, $\sqrt{27}\simeq 5.196$, as denoted by the horizontal line.  The color bar indicates the observer's inclination angle $0^{\circ} \le \theta_{O} \le 180^{\circ}$. There are degeneracies of $\bar{R}$ for different inclination angles, which will then be indicated by the solid lines. Overall, the results show $\frac{|\bar{R}-\bar{R}_0|}{\bar{R}_0}$ is non-zero but less than $0.2\%$.
}
\label{fig:r_avg} 
\end{figure}

It is not clear if there is a change on the average radius $\bar{R}$ of the shadow by judging from the images of Fig. \ref{fig:shadow}.  Thus, we present the numerical result of $\bar{R}$ in Fig. \ref{fig:r_avg} for the three types of soft hair functions given in \eq{CY10}-\eq{CY30}, from which we can read some interesting features. For ${\mathcal E}=0$ we obtain a value $\bar{R}_0\simeq 5.186$, which is consistent with the theoretical value $\sqrt{27}\simeq 5.196$ of the Schwarzschild case within our numerical error bar (indicated by the long vertical line in Fig. \ref{fig:r_avg}). For type I and II soft hair function, the deviation of $\bar{R}$ from $\bar{R}_0$ is still within the numerical error bar, but this is not the case for type III case for large enough $\mathcal E$. Moreover, for a given inclination angle, the average radius $\bar{R}$ can slightly increase or decrease monotonically as $\mathcal E$ increases. In contrast, for a fixed ${\mathcal E}$, we see that $\bar{R}$ may not be monotonically increasing or decreasing as we change the inclination angle. For example, for type I soft hair, we see that $\Delta \bar{R}\equiv \bar{R}-\bar{R}_0$ change from negative values to positive ones as $\theta_{O}$ increase from $0^{\circ}$ to $180^{\circ}$. For type II soft hair, it changes from negative values to positive ones and then back to negative. It alternates between negative values and positive one more time for type III soft hair. Thus, we can conclude that the soft hair will slightly change the average size $\bar{R}$ so that $\frac{\Delta \bar{R}}{\bar{R}_0} \le 0.2 \%$.


From the above results and discussions, we can conclude that the soft hair has a limited effect on the shape of shadow. That is, the soft hair can slightly change the shadow's vertical position and the average radius, but not the horizontal position and its shape deviating from the perfect circle. The shape asymmetry is the most relevant quantity to detect the observational deviation from the Schwarzschild black hole. The other three features will depend on the a priori information about the location and mass/radius of the black hole. This implies that it is difficult to tell the soft hair effects by the shadow unless there are other observational information provided for the location and radius of the black hole. With hindsight, this could be expected because the CLI metric is obtained from the Schwarzschild metric by large gauge transformations. However, the change of radius of the shadow by the soft hair is still surprising, even just by a tiny amount.

\begin{figure}[!ht]
  \centering
  \includegraphics[scale=0.35]{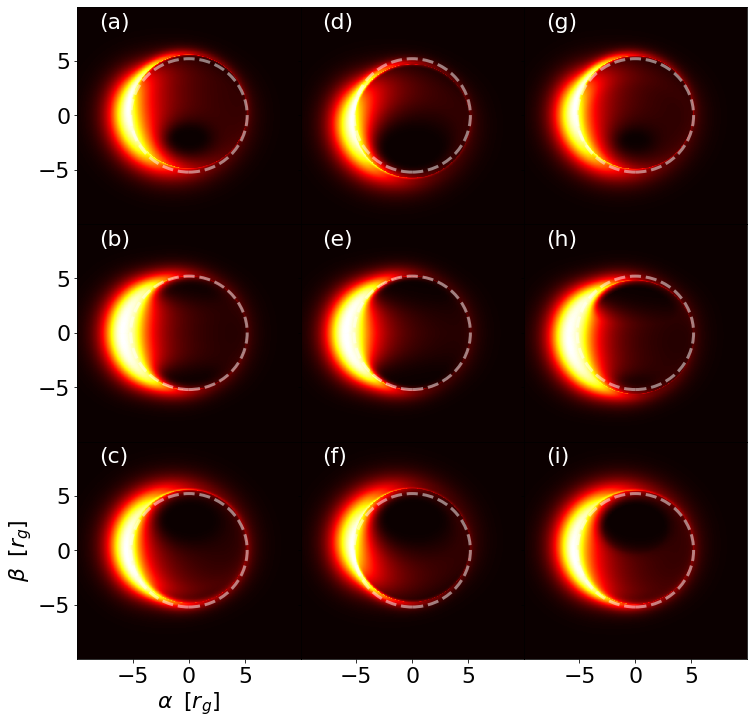}
\caption{Examples of black hole images when supertranslated black holes are surrounded by optically thin and geometrically thick accretion flows. The corresponding soft-hair parameters and inclination angle are the same as in Fig. \ref{fig:shadow}, while the field of view is slightly larger. For reference, the boundary of the shadow for a Schwarzschild black hole is indicated by the dashed white circles. Keeping the same flow dynamics and emission details of the flow, the plots demonstrate that how the resulting image can be affected by detailed radiative transfer process along different geodesics in CLI metrics with different soft-hair perturbations. In each plots, due to the rotation of the accretion flow, the approaching side (left side of each plot) of the flow is brighter than that of the receding side  (right side of each plot).
}
\label{fig:riaf} 
\end{figure}

Images of astrophysical black holes include not only shadow (and therefore the background spacetime) information. The dynamics and spatial distribution of the luminous materials, such as the surrounding accretion flow and jets, would contribute to the  detailed radiative transfer computation along the null geodesics, and therefore the black hole image. 
By considering a phenomenal accretion flow model, we demonstrate the resulting images when a soft-hair black hole is surrounded by an optically thin accretion environment, as presented in Fig. \ref{fig:riaf}.
The details of such Keplerian-rotating, geometrically thick accretion flow is presented in Appendix \ref{appendix-c}. While how the flow dynamics can be affected by the CLI metric is beyond the scope of this paper, here we simply assume the dynamics and material distribution are similar to what can be applied to the Schwarzschild metric, and integrate the radiative transfer contribution (including emission and absorption, see \cite{Pu:2016eml} for example) for thermal synchrotron radiation along the ray.
In  Fig. \ref{fig:riaf}, the same field of view and soft hair parameters as in Fig. \ref{fig:shadow} are adopted. One can therefore compare Fig. \ref{fig:riaf} with Fig. \ref{fig:shadow} to examine how the details of black hole environment can further modify the expected black hole shadow. Although the center of the shadow of supertranslated black holes may be shifted, the shadow would remain spherical, as shown in Fig. \ref{fig:shadow}. Observationally, such a shift would be difficult to identify without a reference center. Nevertheless, minor differences introduced by the different null geodesics can in principle leave imprints on the resulting image. In Fig. \ref{fig:riaf}, plots in each row are of the same inclination angle (but different perturbation condition), and the plots in each column are of the same perturbation condition (but different inclination angles).






\section{Conclusion}\label{sec:5}
In this paper, we studied the shadow of the black hole of soft hairs. Our work is motivated by the recent EHT observations of black hole shadows, and the proposal of black hole soft hairs motivated by the revival interest on the infrared structure of gravity. Our study can be seen as an attempt to explore the black hole information paradox and no-hair theorem from the observational point of view. Our results show that the correction to the shadow image due to soft hair, although tiny, is unique. Namely, within the numerical error, we find that the size and vertical position of the black hole shadow will be modified by the soft hairs, but not the horizontal position and its shape asymmetry deviating from the perfect circle of Schwarzschild case. Moreover, we also show that the accretion patterns of the matter surrounding the black hole will be also affected by the soft hairs.

Our study in this paper is quite preliminary in the observational sense, since we consider only Schwarzschild-like black holes, since the Kerr-like black holes could be more common in the universe. However, at this stage there is no known solution for the Kerr black hole with soft hairs having been constructed. Despite that, the same methodology can be applied once new black hole solutions with soft hairs are found, and the shadow features we have found in this paper can serve as the reference guide for the new constructions. Moreover, the more generic soft hair functions beyond what considered in this paper could introduce more features for the shadow morphology, which requires more intensive study for the future works.

\acknowledgments
We thank Hsu-Wen Chiang and Yu-Hsien Kung for pointing out the neat CLI form, and thank Che-Yu Chen for helpful discussions.  FLL is supported by Taiwan's Ministry of Science and Technology (MoST) through Grant No.~109-2112-M-003-007-MY3. AP is supported by MoST grant No.~109-2811-M-003-505 and No.~110-2811-M-003-507.  HYP is supported by the Ministry of Education (MoE) Yushan Young Scholar Program, MoST grant 110-2112-M-003-007-MY2, and National
Taiwan Normal University. We also thank National Center for Theoretical Sciences' partial support.


\appendix

\section{Geodesic Equations}\label{append-geos}
To obtain the shadow, we need to solve the geodesic equations around the CLI metric, which we just recapitulate here 
\be 
ds^2 = -Vdt^2+\frac{\sec^2\psi}{V}dr^2 + 2\frac{(1-\psi')\tan\psi}{\sqrt{V}}dr d\theta + r^2[(1-\psi')^2 d\theta^2 + \sin^2(\psi-\theta)d\phi^2].
\ee
where 
\bea 
V(r)&=& 1-2M/r, \\
\psi(r,\theta)&=& \sin^{-1}[\frac{2C'(\theta)}{K(r)}], \\
K(r)&=& r-M+\sqrt{r(r-2M)}
\eea
with $':= \frac{d}{d\theta}$. Since the metric is static and polar symmetric, we have the associated constants of motion, i.e., $E$ and $L_z$ which are defined by the following two equations:
\bea
E &\equiv& -\frac{\delta {\mathcal L}}{\delta \dot{t}}=V \dot{t}\;, \label{E-eq} \\
L_z &\equiv& \frac{\delta {\mathcal L}}{\delta \dot{\phi}}= r^2 \sin^2\bigl(\theta -  \psi\bigr) \dot{\phi}\;. \label{L-eq}
\eea

By varying the Lagrangian ${\cal L}={\frac{1}{ 2}} g_{\mu\nu} \dot{x}^{\mu} \dot{x}^{\nu}$ for the light-rays and with the help of \eq{E-eq} and \eq{L-eq}, we can obtain the following geodesic equations:
\bea
\ddot{r} &=& {\frac{1}{2}} \sqrt{V} \sin\bigl(2 \psi\bigr) \bigl(\psi'-1 \bigr) \ddot{\theta} -  2 \dot{r} \dot{\theta} \psi' \tan\psi \partial_\tau r \nn \\ && + \frac{L_{z}{}^2 \cos^2\psi \csc^2\bigl(\theta -  \psi\bigr) \Bigl(\sqrt{V} \cot\bigl(\theta -  \psi\bigr) \tan\psi + V \Bigr)}{r^3} \nn \\&&
- \frac{E^2 \partial_r V \cos^2\psi}{2V} + \frac{\dot{r}^2 \bigl(2 \sqrt{V} \tan^2\psi + r \partial_r V \bigl)}{2 r V}\nn \\
&& + \dot{\theta}^2 \sqrt{V} \biggl(r \sqrt{V}  \bigl(\psi'-1 \bigr)^2 \cos^2\psi  +(1-r) \psi'\bigl(\psi'-1 \bigr)  -  {\frac{1}{2}} \sin\bigl(2 \psi\bigr) \psi'' \biggr)\;, \label{r-geo} \\
\ddot{\theta} &=& \frac{\ddot{r} \tan\psi + 2 r \sec^2\psi \psi' -  \dot{r} \dot{\theta} \biggl(2 r \sqrt{V} \bigl(\psi'-1 \bigr) - 2 r \psi' \sec^2\psi   \biggr)}{r^2 \sqrt{V} \bigl(\psi'-1 \bigr)} \nn \\ 
&& - \frac{L_{z}{}^2 \cot\bigl(\theta -  \psi \bigr) \csc^2\bigl(\theta -  \psi\bigr)}{r^4 \bigl(\psi'-1 \bigr)} -  \frac{\dot{\theta}^2 \psi''}{\bigl(\psi'-1 \bigr)} \nn \\
&& -  \frac{\dot{r}^2 \tan\psi  \biggl(\bigl(\psi'-1 \bigr)r \partial_r V  - 2 \sqrt{V} \bigl(1 + (r-2) \psi'\bigr)\sec^2\psi \biggr)}{2 r^3 V^{3/2} \bigl(\psi'-1 \bigr)^2}\;, \\
\ddot{\phi} &=& - \frac{2 L_{z}{} \csc^2\bigl(\theta -  \psi \bigr) \biggl(\dot{r} \Bigl(\cot\bigl(\theta -  \psi\bigr) \tan\psi + \sqrt{V} \Bigr)  -  r \sqrt{V} \dot{\theta} \bigl(\psi'-1 \bigr)\cot\bigl(\theta -  \psi \bigr) \biggr)}{r^3 \sqrt{V} }\;. \qquad \label{phi-geo}
\eea

Moreover, the ambiguity of choosing the worldline time for a null geodesic can be fixed by imposing the null condition ${\cal L}=0$, which gives
\bea
E^2 &=& \dot{r}^2 \sec^2\psi  + \frac{V  \biggl(L_{z}{}^2 \csc^2\bigl(\theta -  \psi \bigl)+ r^4 \dot{\theta}^2 \bigl(\psi'-1 \bigr)^2 \biggr)}{r^2} - 2 \sqrt{V} \dot{r} \dot{\theta} \bigl(\psi'-1 \bigr) \tan\psi \;. \qquad \label{HJ-eq}
\eea
Once choosing a pixel with a given label $(x,y)$ on the image plane, one can determine the initial conditions for the light-ray in the black hole coordinates through \eq{r0-eq}-\eq{phid0-eq}. With these initial conditions, we can obtain $L_z$ from \eq{L-eq}, and then $E$ from \eq{HJ-eq} even without knowing the relation between the coordinate time $t$ and the worldline time $\tau$. Now we can plug $E$ and $L_z$ into the geodesic equations \eq{r-geo}-\eq{phi-geo}, and then solve them for the light-ray geodesic with the given initial conditions \eq{r0-eq}-\eq{phid0-eq}.

Besides, in section \ref{sec:3} and Appendix \ref{append-a} we will examine if one can analytically obtain the shadow of the black hole with linear supertranslation hair. For this purpose, we expand the above geodesic equations up to the linear order in $C$. The results are (Below, we set $M=1$ for simplicity.)
\bea\label{t-geo-C}
\dot{t}&=&{\frac{E r}{r -2}  }\;,\\
\label{r-geo-C}
0 &=& {r(r-2) \ddot{r} -\dot{r}^2  + E^2 \over (r-2)^2} - r \dot{\theta}^2 - {L_z^2 \csc^2\theta \over r^3} \nn \\
&+&{2 r^2 C' \ddot{\theta} \over \sqrt{r (r-2)} K(r) }+{4 r (2r^2-4r+1+2(r-1)\sqrt{r(r-2)}) \over K(r)^3} C'' \dot{\theta}^2  \nn \\ 
&-& {4(r-2)(3r^2-5r+1)+ 2(6r^2-16r+9) \sqrt{r(r-2)} \over r^3(r-2) K(r)^3} C' L_z^2 \cot\theta \csc^2\theta \;, \\
\label{theta-geo-C}
0 &=& r^2 \ddot{\theta} + 2 r\dot{r} \dot{\theta} - {L_z^2 \cot\theta \csc^2\theta \over r^2} \nn\\
 &-& 2 \Bigg( {L_z^2 (2+\cos2\theta ) \csc^4\theta \over r^2 K(r)}   - {\sqrt{r} \Big( (2r-4-K(r)) \dot{r}^2 +r(r-2) \ddot{r} \Big) \over (r-2)^{3/2} K(r)}\Bigg) C'  \nn \\
 &+& { 2  \Big( 2r^3 (K(r)-3r+5) \dot{r} \dot{r} + (r-2) (L_z^2 \cot\theta \csc^2\theta -2 r^4 \ddot{\theta}) \Big) C'' - r^4 \dot{\theta}^2 C'''\over r^2 (r-2)K(r)}\;, \\
\label{phi-geo-C}
\dot{\phi} &=& L_z \Big( {csc^2\theta \over r^2} + {4 \cot\theta \csc^2\theta C' \over r^2 K(r)} \Big) \;.
\eea 
For completeness, we also expand the null condition ${\cal L}=0$ up to the linear order in $C$ as following. 
\bea
 0 &=&  (L_z^2 \csc^2\theta - {E^2 r^3 \over r-2})+ {r^3 \dot{r}^2 \over r-2} + r^4 \dot{\theta}^2 \nn \\ \label{HJeq-C}
&& + {4 \over (r-1+\sqrt{r(r-2)})}\Bigg(\Big( L_z^2 \cot\theta \csc^2\theta + \sqrt{1\over r(r-2)} r^4 \dot{r} \dot{\theta} \; \Big) C'-r^4 \dot{\theta}^2 C''\Bigg)\;.
\eea 
Note that the first line of each above equation corresponds to the one for the Schwarzschild black hole. We see that even just keeping up to first order in $C$, the geodesic equations are tedious and the explicit expressions are not illuminating.

\subsection{Carter constant and separating  the null condition}\label{app_Carter}

When following the method of \cite{Grenzebach:2014fha} to analytically construct the black hoke shadow, we need to separate the null condition as done in the cases of Schwarzschild and Kerr black holes with the help of existence of Carter constant. For the general metrics, we first need to check if there are additional constants of motion associated with some hidden symmetry such as the Carter constant to proceed the above, and it could be a difficult task. Fortunately, the CLI metric is obtained from the Schwarzschild one, which has the Carter constant $\cal K$, by the following coordinate transformation up to the linear order of $C$,
\be
r_s=r\;, \qquad \theta_s = \theta -{2 \; C'(\theta) \over K(r)}\;. \label{S_2_CLI} 
\ee
so that a scalar quantity such as the Carter constant should be preserved. Here $(r,\theta)$ are the radial and polar coordinates in the CLI metric and $(r_s,\theta_s)$ the ones in the Schwarzschild metric.

Using the above fact, we can obtain the counterparts in the CLI metric from the known ones in the Schwarzschild metric by the coordinate transformation \eq{S_2_CLI}. The latter is given by  
\bea
{\cal K}&=& L_z^2 \csc^2\theta_s + r_s^4 \dot{\theta}_s^2\;, \label{Carter_s} \\
{\cal K}&=&{E^2 r_s^3 \over r_s-2} - {r_s^3 \dot{r}_s^2 \over r_s-2} \label{radial_s}\;,
\eea 
and the resultant CLI counterpart up to ${\cal O}(C)$ is 
\bea 
{\cal K} &=& L_z^2 \csc^2\theta + r^4 \dot{\theta}^2 \nn\\
&& + { 4 L_z^2 \cot\theta \csc^2\theta C'(\theta) \over K(r)}+ {4 r^3 \dot{r}\dot{\theta}  C'(\theta) \over K(r) \sqrt{V(r)} }-{4 r^4 \dot{\theta}^2 C''(\theta) \over K(r)}
\label{Carter_CLI}\;,
\\
{\cal K} &=& {E^2 r^3 \over r-2} - {r^3 \dot{r}^2 \over r-2} \;. \label{radial_CLI}
\eea
We will employ these two equations to construct the shadow analytically.

A side remark is in order regarding the terminology of separability. The usual notion of ``separability" means that we can separate the mixed null condition into two decoupled equations. For example, \eq{radial_s} is already an equation of $r$ only, and we can also turn \eq{Carter_s} into an equation of $\theta$ only by redefining the worldline time from $\tau$ to $\lambda$ (the so-called Mino time) by $d\lambda={d\tau \over r^2_s}$. However, it is easy to see that it is impossible to make \eq{Carter_CLI} an equation of $\theta$ only in the same way. This just reflects the fact that separability depends on the coordinates employed. As emphasized, for the analytical construction of the shadow, we just need to separate the null condition into two separated ones with the help of Carter constant even though they are not decoupled.

\section{Analytical Approach to solve the shadow of black hole with linear supertranslation hair}\label{append-a}
In this appendix, we attempt to analytically construct the shadow of a black hole with linear supertranslation hair by following the method of \cite{Grenzebach:2014fha}. 

As shown in Fig. \ref{fig:schematic}, the point $(X,Y)$ on the celestial image plane is related to the initial tangent vector
\be\label{lambda1}
\dot{\lambda}:= \dot{t}\partial_t + \dot{r}\partial_r + \dot{\theta} \partial_{\theta} + \dot{\phi} \partial_{\phi}
\ee
at observer point $(r_O, \theta_O)$ in the CLI coordinate by
\be \label{sdmap}
X=-2 \tan{\Theta\over 2} \sin\Psi\;, \qquad Y=-2 \tan{\Theta\over 2} \cos\Psi 
\ee
where $(\Theta,\Psi)$ are the angular coordinates labelling  $\dot{\lambda}$ in the observer's local frame, i.e.,
\be \label{lambda2}
\dot{\lambda}= -{E\over \sqrt{V(r)}} (-e_0 + e_1 \sin \Theta \cos \Psi + e_2 \sin \Theta \sin \Psi + e_3 \cos\Theta )\;.
\ee
The orthonormal tetrads around the observation point based on CLI metric \eq{CLI} are given by
\begin{align*}
    e_0 &= \left.\frac{1}{\sqrt{V(r)}}\partial_t\right|_{(r_O,\theta_O)}\;,\\
    e_1 &= -\frac{2 \sqrt{V(r)} C'}{K(r)}\partial_r+\left.\frac{1}{r}\left(1+\frac{2C''}{K(r)}\right)\partial_{\theta}\right|_{(r_O,\theta_O)}\;,\\
    e_2 &= \left.-\frac{1}{r\sin{\theta}}\left(1+\frac{2 C' \cot{\theta}}{K(r)}\right)\partial_{\phi}\right|_{(r_O,\theta_O)}\;,\\
    e_3 &= \left.-\sqrt{V(r)}\partial_r\right|_{(r_O,\theta_O)}\;.
\end{align*}
Matching \eq{lambda1} to \eq{lambda2} we obtain the following relations
\bea 
\dot{r} &=& E \Big(\cos\Theta + {2 \sin\Theta \cos\Psi \; C'(\theta_O) \over r_O -1+ \sqrt{r_O (r_O -2)}}  \Big)\;, \label{nray-r} \\
\dot{\theta} &=& - {E \sin\Theta \cos\Psi \over \sqrt{r_O(r_O -2)}} \Big(1 + {2 C''(\theta_O) \over r_O-1+ \sqrt{r_O (r_O -2)}}  \Big) \;, \label{nray-theta} \\
\dot{\phi} &=& {E \csc\theta_O \sin\Theta \sin\Psi \over \sqrt{r_O(r_O -2)}} \Big(1+ {2\cot\theta_O \; C'(\theta_O) \over r_O-1+\sqrt{r_O(r_O -2)}} \Big) \label{nray-phi} \;.
\eea
Using the geodesic equations \eq{t-geo-C}-\eq{HJeq-C}, we can try to solve $(\Theta,\Psi)$ in terms of $(r_O,\theta_O)$ and the conserved quantities. By further plugging the conserved quantities for the photon sphere into the above solution, we can obtain $(\Theta,\Psi)$ or $(X,Y)$ of the shadow on the celestial image plane. 

To proceed with the above matching and solving steps in the perturbative expansion of $C$, i.e.,
\be \label{ThPs1} 
\Theta= \Theta_0 + \Theta_1\;, \qquad \Psi=\Psi_0 + \Psi_1
\ee 
where the subscripts $0$ and $1$ denote the leading and first order results, respectively. At the leading order we obtain the known result for the Schwarzschild black hole, 
\be \label{shadow0}
\sin\Theta_0 = \sqrt{ 27 (r_O-2) \over  r_O^3 }\;, \qquad 
\sin\Psi_0 = {\csc\theta_O \over 3\sqrt{3}} b_0
\ee 
where we define the impact parameter
\be
b_0:={L_z \over E}\;.
\ee
Thus, as we change $b_0$, the light-ray delineates the shape of shadow on the celestial image plane. From \eq{shadow0} and \eq{sdmap} it is easy to see that the shadow is a perfect circle of radius $r_{shadow}=2 \tan\Big({1\over 2}\sin^{-1}(\sqrt{ 27 (r_O-2) \over  r_O^3 })\Big)$.

Note that in arriving the above results, we have taken the advantage of the separability of the geodesic equations because of the full spherical symmetry of background spacetime. Once the soft hair is included, the polar symmetry gets lost, so that the null condition \eq{HJeq-C} can no longer be separated as the case of the zeroth order. Despite that, in Appendix \ref{app_Carter}, by coordinate transformation from the Schwarzschild metric to the CLI metric, we can obtain two separate but non-decoupled equations. Overall, we have three first-order equations of motion for the light-ray geodesic, namely, \eq{phi-geo-C}, \eq{Carter_CLI}, and \eq{radial_CLI}.

To obtain the equations for solving $(\Theta_1,\Psi_1)$, we plug \eq{nray-phi} into \eq{phi-geo-C}, and also plug \eq{nray-r} and \eq{nray-theta} into \eq{Carter_CLI}, and \eq{radial_CLI}. As a consistency check, it is straightforward to show that only twos of the three resultant equations are linearly independent. The one from \eq{phi-geo-C} gives (with $M=1$ and ${\cal K}=27 E^2$)
\be\label{phi-shadow} 
\Theta_1 b_0 \sqrt{{r_O^2 \over 27 V(r_O)}-1}  - \Psi_1  \sqrt{27-b_0^2 \csc^2\theta_O} \sin\theta_O = {2 b_0  C'(\theta_O) \cot\theta_O \over K(r_O)}\;,
\ee
and the other linearly independent one from \eq{radial_CLI} gives 
\be \label{Theta1}
\Theta_1 = -{2 C'(\theta_O) \over K(r_O)} \sqrt{1-{b_0^2 \csc^2\theta_O \over 27}}\;.
\ee
From the above, we can solve $\Psi_1$ to be
\be
\label{Psi1}
\Psi_1 = -{2 b_0 \csc\theta_O C'(\theta_O) \over 27 K(r_O)} \Bigg( {27 \cot\theta_O \over \sqrt{27 - b_0^2 \csc^2\theta_O}} + \sqrt{{r_O^2 \over V(r_O)}-27} \Bigg)\;. 
\ee
Plugging \eq{ThPs1}, \eq{shadow0}, \eq{Theta1} and \eq{Psi1} into \eq{sdmap}, we can obtain the shadow of the black hole with soft hair $C(\theta)$ by varying the impact parameter $b_0$ for a given observer coordinate $(r_O,\theta_O)$. Some typical results will be shown in the main text.

\section{Phenomenological accretion flow model}\label{appendix-c}
Here we describe the details of the accretion flow model mentioned in section \ref{sec:4}.
When the black hole accretion rate is low enough, such as the cases of the two primary targets of the EHT: M87$^{\star}$ and Sgr A$^{\star}$, the accretion flow is radiatively inefficient as the hot ions in the flow cannot efficiently transfer their heat to the electrons.  We consider an accretion flow of such radiatively inefficient accretion flow (RIAF) and follow a phenomenological accretion flow model described in \cite{bro06}:
the electron temperature $T_{e}$ and electron density $n_{e}$ are described by 
\begin{eqnarray}
n_{e}   &=& n^{0}_{e,\rm th}\;\,r^{-\alpha} e^{-z^{2}/(2\rho^{2})}\;, \label{eq:p1} \\
T_{e}             &=& T_{e}^{0}\;\,r^{-\gamma}\;, \label{eq:p2} 
\end{eqnarray}
where $z\equiv r\cos\theta$ and $\rho\equiv r\sin\theta$. The flow dynamics is assumed to be Keplerian rotating outside the inner most stable circular orbit (ISCO) of the Schwarzschild metric, and free-fall within ISCO. Such semi-analytical model has wide applications in, e.g., \cite{bro11,bro14,psa15,pu18}.

Assuming the magnetic field  strength is approximately equipartition with the ions and the flow dynamics can be approximated by that of Schwarzschild case, we  compute the resulting thermal synchrotron image when the soft-hair black hole is surrounded by the aforementioned RIAF environment and the observation frequency $f_{\rm obs}=230$GHz, the observation frequency for current EHT observation \cite{Akiyama:2019cqa,Akiyama:2019brx,Akiyama:2019sww,Akiyama:2019bqs,Akiyama:2019fyp,Akiyama:2019eap}. The distance to the black hole ($\sim8.2$ kpc) and the mass of the black hole ($\sim4.2\times10^6 M_{\odot}$) are assumed to be similar as that for Sgr A$^{\star}$ \cite{gravity19}. The
parameters $\alpha=1.1$, $\gamma=0.84$, $n_{e}^{0}=10^7$ cm$^{-3}$, and $T_{e}^{0}=1.7\times 10^{11}$ $K$ are adopted, similar to that considered in \cite{bro06}. 

To construct the 4-velocity of the flow in the CLI metric, $u^{\alpha}=(u^{t},u^{r},u^{\theta}=0,u^{\phi})$, we assume the radial component, $u^{r}$, and the orbital frequency, $u^{\phi}/u^{t}$, are the same as that of a Keplerian rotating flow in the Schwarzschild case. By applying $u^{\alpha}u_{\alpha}=-1$, the flow dynamics can be uniquely specified \cite{pu16_flow}. With all the setup introduced above, the resulting total flux of the computed images in Fig. \ref{fig:riaf} is $\sim1.5-2.5$ Jy, similar to that of Sgr A$^{\star}$ (e.g. \cite{bow18}).


\bibliographystyle{unsrt}
\bibliography{references.bib} 

\end{document}